\documentclass[]{article}
\usepackage{authblk}
\pdfoutput=1

\usepackage[utf8]{inputenc} 
\usepackage[T1]{fontenc}    
\usepackage{hyperref}       
\usepackage{url}            
\usepackage{booktabs}       
\usepackage{amsfonts}       
\usepackage{nicefrac}       
\usepackage{microtype}      
\usepackage{fullpage}

\usepackage[export]{adjustbox}
\usepackage{hyperref}
\usepackage{url}
\usepackage{graphicx}
\usepackage{amsmath}
\usepackage{bbm}
\usepackage{paralist}
\usepackage{Definitions}
\usepackage[ruled,noline]{algorithm2e} 
\usepackage{color}
\usepackage{enumitem}
\usepackage{mathtools}
\usepackage{subfig}
\usepackage{graphicx}
\usepackage{tabularx}
\usepackage[dvipsnames]{xcolor}

\newcommand{\xhdr}[1]{\vspace{1mm}\noindent{{\bf #1. }}}

\newcommand{\badge}[1]{\ifmmode\text{\texttt{#1}}\else\texttt{#1}\fi}

\newcommand{\explain}[2]{\underset{\mathclap{\overset{\uparrow}{#2}}}{#1}}
\newcommand{\explainup}[2]{\overset{\mathclap{\underset{\downarrow}{#2}}}{#1}}
\newcommand*{\defeq}{\coloneqq}

\title{Determining Impact of Social Media Badges through\\ Joint Clustering of Temporal Traces and User Features}

\author[1]{Tomasz Kusmierczyk}
\author[2]{Kjetil N{\o}rv{\aa}g}

\affil[1,2]{Norwegian University of Science and Technology, \{tomaszku,noervaag\}@idi.ntnu.no}

\date{}

\begin{document}


\maketitle

\begin{abstract}
Badges are a common, and sometimes the only, method of
incentivizing users to perform certain actions on online sites.
However, due to 
many competing factors influencing 
user temporal dynamics,
it is
difficult to determine
whether 
the badge had (or will have) the intended effect or not.

In this paper,
we introduce 
two
complementary
approaches 
for
determining
badge influence on users.
In the first one,
we cluster 
users' temporal traces 
(represented with point processes)
and apply covariates (user features) to regularize results.
In the second approach, we first  classify users'
temporal traces with a novel statistical framework, and then we refine the classification results with a semi-supervised clustering of covariates.

Outcomes obtained 
from an evaluation on synthetic datasets
and
experiments
on two
badges from a popular Q\&A platform
confirm that 
it is possible to 
validate, characterize
and to some extent predict 
users affected by the badge.
\end{abstract}

\section{Introduction}
\label{sec:intro}
Awarding a digital badge after a user performs certain actions is a
common mechanism to motivate users on online sites, be it social
networking sites like
Foursquare\footnote{\url{https://foursquare.com}}, education sites like
Khan Academy\footnote{\url{https:// www.khanacademy.org/}}, or
crowdlearning Q\&A sites like Stack
Overflow\footnote{\url{https://stackoverflow.com}}.  Previously, there
have been several attempts at modeling the badges' effect on online
communities and at recommending how the badge systems should be
designed.  However, there are no previous studies actually verifying whether the
badges have any impact on individual users or not; it has been taken
for granted that badges affect targeted users in a desired way.  In
contrast, in this paper we take a closer look at this assumption, and present
the first work that addresses the problems of \textit{validation,
characterization and prediction of users attracted to badges}. 
\begin{figure}[t!]
\centering
\includegraphics[width=0.35\textwidth,valign=c]{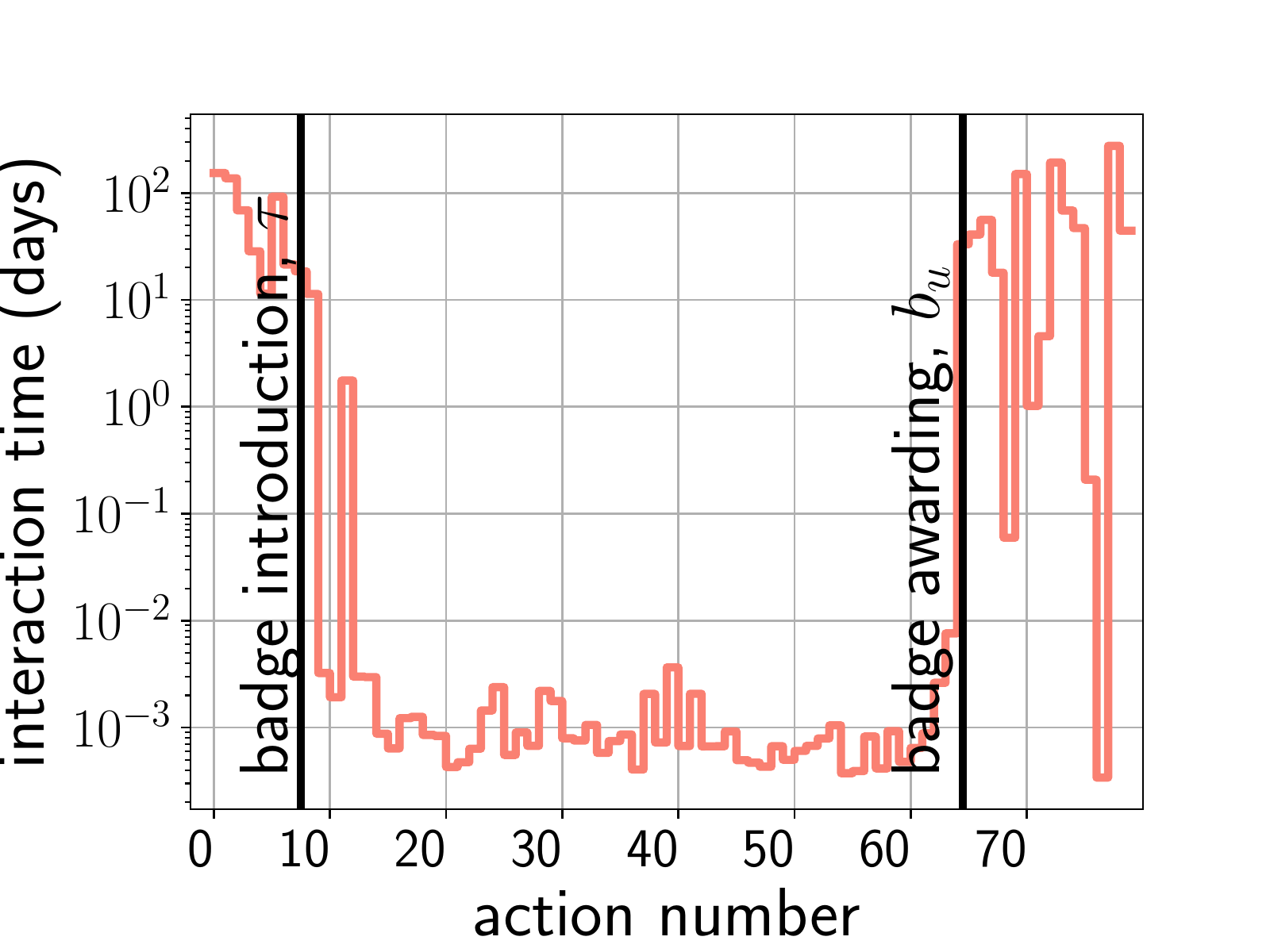}
\includegraphics[width=0.225\textwidth,valign=c]{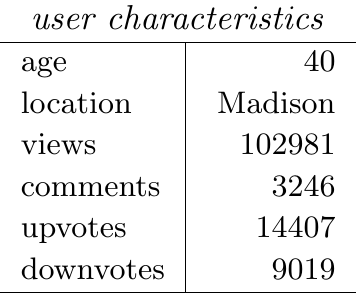}
\caption{Sample user from Stack Overflow influenced by
the \badge{Research Assistant} badge that is awarded for tag wiki
edits. The user increases its action rate when the badge is introduced to the community (at time $\tau$), and returns to the previous rate after receiving it (at time $b_u$).}
\label{fig:sample_user}
\end{figure}

It is a challenging task to answer the question whether a user was (is) in
any way motivated by the badge.
Users 
tend to evolve over time,
and
apart from badges
there are usually many competing factors 
influencing their dynamics.
Furthermore, neither ground truth 
nor 
counterfactual data
showing behavior of users not influenced by badges 
are available.

In this paper,
we focus on the most popular type of badges,
\ie, \textit{threshold badges}, 
that are awarded after a user performs 
a certain number of desired actions.
The above challenges 
in this context
can be addressed
by simultaneously looking at 
users' temporal traces 
and their general characteristics.
In particular,
we identified
the following
useful
patterns: 
\begin{itemize}[noitemsep,nolistsep,leftmargin=0.7cm]
\item[--] attracted users change their \emph{mean} behavior around the badge awarding time 
\item[--] influenceable users are similar 
\end{itemize}
As an example, Figure~\ref{fig:sample_user} illustrates 
how user action rate changes due to the badge, 
and also shows the associated
user features and statistics.
In this paper, we propose 
two complementary solutions to
the 
\textit{user badge influence problem}
exploiting 
the above 
observations.
In both
we apply users' 
temporal traces
(modeled as \textit{non-homogeneous Poisson processes})
as a main source of the badge effect information
and use 
associated covariates (user features and statistics)
to regularize classification results. 

\xhdr{Summary of Contributions} 
This paper makes the following contributions:
\begin{itemize}[noitemsep,nolistsep,leftmargin=0.5cm]
\item[--] introduction and formalization of the \textit{user badge influence problem} 
\item[--] validation and prediction of the influence of badges on
individual users with two novel methods:
	\begin{itemize}[noitemsep,nolistsep,leftmargin=0.5cm]
	\item[I.] 
	a model-based algorithm 
	to cluster (problem-specific) counting processes with covariates used to encode priors
	\item[II.] 
	a statistical test complemented with a way of calibrating it
	by means of \emph{virtual badges bootstrapping} and an
	adaptation of the EM clustering algorithm that refines the results of this test
	\end{itemize}
\item[--] empirical evaluation using synthetic data 
\item[--] case studies of two badges from a popular Q\&A platform
\end{itemize}
In the rest of the paper we'll give an overview of related work,
formalize the problem, provide a detailed description of  our
proposed solutions, and demonstrate their effectiveness on synthetic
and real datasets.
\section{Related Works}
\label{sec:related}

Our work extends previous studies on
motivational mechanisms in social media~\cite{ghosh2011incentivizing,hamari2014does,lewis2004influence}
and in particular
on understanding and modeling the effects of badges~\cite{anderson2013steering,gibson2015digital,zhang2016social}.

Previously,
\cite{ImmStoSyr15,easley2016incentives,zhang2016social} 
worked on 
optimal badges design from a game-theoretic perspective. 
They relied on strong theoretical assumptions not necessarily satisfied in real data,
including the assumption that badges \emph{always} work.
In this paper, we challenge this presumption.
Early studies suggesting that may not always be the case appeared first 
in the educational context~\cite{abramovich2013badges}.
In the context of social media the problem of badges effectiveness was noticed very recently~\cite{FM7299,1707.08160,HAMARI2017469}. 
In these works researchers assumed badges effect to be binary, \ie, either a badge changes community functioning or not,
and 
(apart from \cite{1707.08160} who on the other hand worked with simpler \textit{single-action badges})
focused on site-global statistics (total number of views, edits, etc.),
whereas we take a user-level perspective and try to understand badge influence on individuals.
This places our study closer to works trying to characterize susceptible users in social media, for example \cite{aral2012identifying},
and modeling user behavior in presence of badges~\cite{anderson2013steering,mutter2014behavioral}.
The latter two rely on a \emph{goal-gradient hypothesis} (users become more active closer to badge)
that 
we found hard
to observe in individual users traces.
In contrast, we focused on mean changes in user behavior around the
time of badge awarding. 
\section{Problem Formulation} 
\label{sec:problem}

In this section, we present the problem of 
\textit{determination of badge influence on a user}
in a formal way,
and
introduce 
a point process model of user behavior 
in context of a badge.

\xhdr{Notation: Badges and Users} 
A digital \textit{threshold badge} $b$ can be represented with a tuple:
\begin{equation*}
b~~\defeq~~(\explain{\tau}{\text{introduction}}, \quad \explainup{a}{\text{desired actions type}}, \quad \explain{T}{\text{threshold}}), 
\end{equation*}
where $\tau$ is the badge introduction time (=~the time when the badge started being awarded), $a$ is an assigned action type, and
$T$ is the badge threshold (=~the number of type $a$ actions that need
to be performed by a user in order to be awarded the badge). 

User $u \in U$ 
in context of the badge $b$
can be represented 
by a tuple: 
\begin{equation*}
u~~\defeq~~(\explain{s_u}{\text{start time}}, \quad \explainup{e_u}{\text{end time}}, \quad \explain{{{\vec{x_u}}}}{\text{user features}}~, \quad \explainup{\{t_u\}}{\text{action times}}, \quad \explain{{{b_u}}}{\text{badge awarding}}~, \quad \explainup{{\color{gray} i_u}}{\text{badge attraction}}), 
\end{equation*}
where $s_u$ and $e_u$ designate user activeness interval (time span in which we test the badge effect),
$\vec{x_u}$ is a vector of badge covariates (\eg, user characteristics),
$\{t_u\}$ is a set of timestamps of desired (=type $a$) actions,
$b_u$ is the time when user $u$ received badge $b$
(=achieved level of $T$ actions).
If the user $u$ has not received the badge yet
(\ie, ~ $ |\{t_u\}|< T$), we set $b_u = \infty$.
Finally, the binary variable ${\color{gray} i_u}$ informs if the user is/was attracted by the badge reward perspective or not
(the fact that user received a badge does not necessary imply that she had any interest in that -- it could be just a side-effect of her normal activity).

Additionally, to simplify some of the later formulations, we define:
$l_u=e_u-s_u$, 
$l^0_u=b_u-s_u$, 
$l^1_u=e_u-b_u$,
$n_u=|\{t_u\}|$,
$n^0_u=|\{t_u:~t_u<b_u\}|$, 
$n^1_u=|\{t_u:~t_u~\geq~b_u\}|$.

\xhdr{Influenced Users Validation and Prediction} 
We distinguish users influenced by the badge $b$ from those not attracted  
via 
the binary variable ${\color{gray} i_u}$. 
Unfortunately, 
the variable is usually 
hidden. 
Its value recovery 
can be done in two practical settings:
\begin{itemize} [noitemsep,nolistsep,leftmargin=0.7cm]
\item[I.] \textit{Validation}: user received the badge ($b_u~<~\infty$) and we verify if it did not happen just by chance.
\item[II.] \textit{Prediction}: user has not received the badge yet ($b_u~=~\infty$) and we try to forecast if she may be interested in receiving it.
\end{itemize}

For neither of the tasks we know the truth. 
Therefore, we rely only on our assumptions relating
badge influence with
temporal traces ($\{t_u\}$)
and 
users' general characteristics (encoded in $\vec{x_u}$).

\xhdr{Temporal Traces Model} 
It might be hard to observe if
users attracted by the badge change their behavior when they receive it.
However, temporal fluctuations and impact of
competing factors can be 
reduced with averaging over time.
In particular,
we assume the following model 
of underlying temporal traces,
where
user $u$'s 
action times 
are 
drawn from the \textit{point process}~\cite{daley2002introduction} 
controlled by the intensity $\lambda_u(t)$ that takes one of the two forms: $\lambda^0_u(t)$ or $\lambda^1_u(t)$, depending on the latent variable $i_u$:
\begin{itemize}[noitemsep,nolistsep,leftmargin=0.7cm]
\item[--] $i_u=0$ (user not attracted by the badge): intensity is a constant (user does not change her behavior over time)
\item[--] $i_u=1$ (user attracted by the badge): actions mean intensity changes when the badge is awarded at $b_u$
\end{itemize}
However the our approach was inspired by the ideas from~\cite{1707.08160}, 
we model every individual user with a separate \textit{non-homogeneous Poisson process}   
whereas 
what they proposed is a
\textit{survival process} of all users considered together.

Formally, the model is expressed as follows:
{
\begin{align}
\label{eq:basic-model1}
\{t_u\} & \sim PP(\lambda^{i_u}_u(t)) \nonumber \\
\lambda^0_u(t) &=
\begin{cases}
  0 & \text{if}\enskip t < s_u \lor t > e_u\\
  {\lambda^0(u)} & \text{otherwise}
\end{cases} \\
\lambda^1_u(t) &=
\begin{cases}
  0 & \text{if}\enskip t < s_u \lor t > e_u \\
  {\lambda^1_0(u)} & \text{if}\enskip s_u < t \leq b_u\\
  {\lambda^1_1(u)} & \text{otherwise}
\end{cases}  \nonumber 
\end{align}
}

\section{Learning Attracted Users via Poisson Processes Clustering}
\label{sec:bayesian}

In this section, we introduce a novel model-based algorithm to cluster
Poisson processes,  that we use to identify users influenced by the
badge.  Its extended version employs covariates to regularize clusters
assignment priors and allows for new users prediction.

\xhdr{Basic Model}
We assume
that 
the fraction of users 
attracted by the badge
(having $i_u=1$)  
is $\pi$,
and
the intensities $\lambda$ (expressed in Eq.~\ref{eq:basic-model1})
come from the shared prior gamma distributions:
{
\begin{align}
\label{eq:robust-survival-model}
i_u & \sim \mbox{Bernoulli}(\pi) \nonumber \\
\lambda^0(u) &\sim \mbox{Gamma}(\alpha^0, \beta^0) \nonumber \\
\lambda^1_0(u) &\sim \mbox{Gamma}(\alpha^1_0, \beta^1_0)  \\
\lambda^1_1(u) &\sim \mbox{Gamma}(\alpha^1_1, \beta^1_1) \nonumber
\end{align}
}
The full model then has seven hyperparameters:
$\theta^0 = \{\alpha^0, \beta^0\}$, $\theta^1=\{\alpha_0^1, \beta_0^1, \alpha_1^1, \beta_1^1\}$ 
 steering the behavior of users with respectively $i_u=0$ and $i_u=1$,
 and $\pi$ controlling the fraction of users attracted by the badge. 
Latent variables are  $i_u$ and
action intensities $\lambda^0(u)$ and  $\{\lambda^1_0(u), \lambda^1_1(u)\}$. 

The model can be factorized thanks to independence between user probabilities and independence between Poisson processes on non-overlapping intervals, 
and then simplified via 
marginalization
of latent intensities.
The procedure leads to the following conditional user probabilities:
{
\begin{align}
\label{eq:user_probs}
P(\{t_u\} | \theta_0, i_u=0) =  & \frac{{\beta^0}^{\alpha^0}}{(l_u+\beta^0)^{\alpha^0+n_u}} \frac{\Gamma(\alpha^0+n_u)}{\Gamma(\alpha^0)}  \nonumber
\\
P(\{t_u\} | \theta_1, i_u=1) = & \frac{{\beta^1_0}^{\alpha^1_0}}{(l^0_u+\beta^1_0)^{\alpha^1_0+n^0_u}} \frac{\Gamma(\alpha^1_0+n^0_u)}{\Gamma(\alpha^1_0)} \cdot  
\\
 & \cdot \frac{{\beta^1_1}^{\alpha^1_1}}{(l^1_u+\beta^1_1)^{\alpha^1_1+n^1_u}} \frac{\Gamma(\alpha^1_1+n^1_u)}{\Gamma(\alpha^1_1)} \nonumber
\end{align}
}
The model collapses to a mixture-model with two clusters determined by
$i_u=0$ and $i_u=1$ and controlled by hyperparameters $\theta^0$ and
$\theta^1$, and with mixing factor $\pi$. Cluster assignments and
hyperparameters in this class of models are typically inferred with
an \textit{EM-like procedure} that consists of two alternating steps 
taking in our case the following form:

\begin{itemize}[noitemsep,nolistsep,leftmargin=0.7cm]
\item[I.] \textit{Maximization}: hyperparameters are updated: 
{

$$
\left \{     
\begin{matrix}
\theta^0_{new} \\ \theta^1_{new} \\ \pi_{new} 
\end{matrix}
\right \} 
= \argmax_{\theta^0, \theta^1, \pi} \sum_u \log P(\{t_u\}, i_u | \theta^0, \theta^1, \pi) 
$$
}
where the complete-data likelihood per user relies on per-cluster user likelihoods expressed in Eq. \ref{eq:user_probs}:
{

\begin{align*}
\log & P(\{t_u\}, i_u  | \theta^0, \theta^1, \pi) =  \\
 &  \gamma(i_u) ( \log P(\{t_u\} | \theta^1, i_u=1) + \log \pi ) + \nonumber \\
 &  (1-\gamma(i_u)) ( \log P(\{t_u\} | \theta^0, i_u=0) + \log (1-\pi) ) \nonumber
\end{align*}
}
A closed-form solution to the optimization
problem does not exist.
Instead, we first find cluster probabilities: 
{

\begin{align}
\label{eq:user_pi}
\pi_{new} = \frac{\sum_{u \in U} \gamma(i_u)}{|U|}
\end{align}
}
and then resort to numerical optimization with positivity constraints to find $\theta^0_{new}$ and $\theta^1_{new}$
\item[II.] \textit{Expectation}: posterior cluster responsibilities are found in the usual way:
{

\begin{align*}
\gamma(i_u) =  \frac{ P(\{t_u\} | \theta^1, i_u=1) \pi } {P(\{t_u\} | \theta^1, i_u=1) \pi + P(\{t_u\} | \theta^0, i_u=0) (1-\pi) }
\end{align*}
}
\end{itemize}

\xhdr{Including Covariates}
Badges attract users of similar characteristics
and
therefore user influence covariates $\vec{x_u}$
can be applied
for clustering improvement as a form of regularization.
We incorporate them in the our hierarchical model
similar to  \cite{user-exposure} by replacing
the constant cluster membership prior with user personalized ones, 
\ie, $\pi \rightarrow \pi_u$ that
we furthermore posit to have a functional form:
{

\begin{align}
\pi_u = f(\vec{x_u}, \vec{w}) \in [0, 1]
\label{eq:functional}
\end{align}
}
where $\vec{w}$ are parameters of the function $f$.
In general, $f$ can be any function (for example neural network)
but due to its simplicity we choose logistic regression, 
\ie, $f(\vec{x_u}, \vec{w})~=~\text{sigmoid}(\vec{w}\cdot\vec{x_u})$.

Conditional independence between the priors for cluster memberships and 
clusters' parameters 
implies that the 
inference procedure described above
can be adjusted in a simple way by 
replacing the updates in Eq.~\ref{eq:user_pi} with 
the following optimization of 
vector $\vec{w}$:
{

$$\vec{w}_{new} = \argmax_{\vec{w}} \sum_u (f(\vec{x_u}, \vec{w}) - \gamma(i_u))^2$$
}

\xhdr{Prediction} For a new user without temporal trace 
we predict badge attraction 
only relying on her features and statistics: 
{

$$\hat{P}(i_u) = f(\vec{x_u}, \vec{w})$$
}

\section{Learning Attracted Users with NHST and Covariates Clustering}
\label{sec:frequentist}

In this section, we propose a two-phase procedure validating badge influence on users. In the first phase, we approximately identify users influenced by the badge with a robust \textit{Null Hypothesis Significance Testing} (NHST) procedure.
In the second phase, 
we refine assignments 
with a semi-supervised clustering of covariates.

\subsection{Robust Validation of Attracted Users}

\xhdr{Behavior Change Testing}
The alternative 
that a user $u$ was or was not attracted by the badge can be 
 expressed in terms of  
the null and the alternative hypotheses:
\begin{align*}
H_0 & \,:\, i_u = 0~(\textit{badge $b$ did not have an effect on user $u$}) \\
H_1 & \,:\, i_u = 1~(\textit{badge $b$ influenced user $u$})
\end{align*}
Under the model in Eq~\ref{eq:basic-model1} 
we can restate it in the following way: 
\begin{align*}
H_0 & \,:\, \lambda^{i_u}_u = \lambda^0_u(t) \\
H_1 & \,:\, \lambda^{i_u}_u = \lambda^1_u(t)
\end{align*}

\xhdr{Test Statistic} 
We use a standard log-likelihood ratio
between likelihoods corresponding to $H_0$ and $H_1$
as a test statistic~\cite{hogg1995introduction} which in our case takes the following form:
\begin{align*}
LLR(\lambda^0(u), \lambda^1_0(u), \lambda^1_1(u)) = 
  n_u  \log \lambda^0(u) - l_u \lambda^0(u) + 
- n^0_u  \log \lambda_0^1(u) + l^0_u \lambda_0^1(u) 
- n^1_u  \log \lambda_1^1(u) + l^1_u \lambda_1^1(u) 
\end{align*}
where we plug-in MLE estimates for respective intensities:
$\hat{\lambda^0}(u)=\frac{n_u}{l_u}$,
$\hat{\lambda_0^1}(u)=\frac{n^0_u}{l^0_u}$,
$\hat{\lambda_1^1}(u)=\frac{n^1_u}{l^1_u}$
and assume that $(0 \log 0) = 0$.

\xhdr{Robust Estimation of the Test Statistic Distribution} 
Asymptotically the test statistic $-2 LLR$
for \textit{nested models}
has an approximate chi-square distribution \cite{wilks1938large} with the number of degrees of freedom equal to difference between compared models, \eg, in our case $df=1$. 
The test statistic transformation to p-value is then given by: $p \approx 1-\boldsymbol{\chi}^2_{df}(-2 LLR)$ where $\boldsymbol{\chi}^2$ is a chi-square CDF. 

The standard procedure can detect a change in user behavior happening around $b_u$, but
is not able to differentiate between 
the badge causal effect 
and 
other competing factors.
Instead,  
we design and apply the calibration procedure 
(similar to \cite{1707.08160})
that accounts for them by simulating a counter-factual world where the badge was never awarded and measuring the strength of observed changes there. In practice the test statistic empirical distribution 
is estimated
with the following \textit{virtual badges bootstrapping} procedure: 
\begin{enumerate}[noitemsep,nolistsep,leftmargin=0.7cm]
\item Sample $B$ \textit{virtual badges} $b'_u \sim U([s_u, b_u-m] \cup [b_u+m, e_u])$ where $m$ is some small margin. 
\item Remove the true badge effect by putting it outside the updated activeness limits:
$$(s'_u,~e'_u)~=~
\begin{cases}
(s_u, b_u-m) & \mbox{if}~b'_u<b_u\\
(b_u+m, e_u) & otherwise 
\end{cases}$$
\item Evaluate $LLR'$ with simulated $b'_u, s'_u, e'_u$ and adequately updated $\{t'_u\}$.
\item Approximate empirical p-value: $p~=~\frac{|\{LLR'>LLR\}|}{B}$
\end{enumerate}
\subsection{Assignment Refining via Semi-supervised Clustering of Covariates}

\xhdr{NHST Assignments Misclassification}
The above testing procedure applied to each user splits the population into two groups: 
positives ${P}$ for whom we managed to reject $H_0$ at significance level $\alpha$ and 
negatives ${N}$ for whom we failed to reject $H_0$. 
Although 
this can be used as a first approximation to 
$i_u$,
both groups contain many misclassified cases.
In particular, the
\textit{false positives rate} (\textit{FPR}) and the \textit{false negatives rate} (\textit{FNR}) depend on the 
\textit{statistical test power} and \textit{prevalence} of the
positives over negatives, 
that both are unknown. 
For example, \cite{Sellke2001,Colquhoun140216} estimate \textit{FPR} to be at least around 25\%
when the prior probability of a real effect is $0.5$ and $\alpha=0.05$.
This  means that at least $1/4$ of users initially assigned ${i_u}=1$ actually have $i_u=0$.
For ${i_u}=0$, the fraction of misclassified cases would be even higher.

\xhdr{Semi-Supervised Clustering with Group Priors}
We achieve the reduction of the above classification error
employing a novel 
semi-supervised extension to the 
standard \textit{EM algorithm for gaussian mixtures}~\cite{bishop}
The extended algorithm 
works (=clusters users)
in covariates space 
but additionally employs the 
information transferred from the first (=NHST) phase.
In particular,
initial user assignments and our beliefs about misclassification rates
we encode 
in priors to cluster assignments (=mixing coefficients).

NHST classification splits users into two groups, where ${P}$ and
${N}$ are respectively users initially classified as positives and
negatives.  For each group $G \in \{P, N\}$ we propose to use
separate mixing coefficients $\vec{\pi_G}$
with
Dirichlet hyperpriors, \ie,
$\vec{\pi_G}~\sim~\text{Dirichlet}(\alpha_G^0, ..., \alpha_G^K)$,
where $K$ is a standard parameter controlling the number of clusters
(in contrast to Poisson processes clustering, in covariates space we
can have arbitrary number of clusters).  
In order to be able to interpret clustering results, for each cluster we assign either $i_u=1$  (clusters denoted as $C^1$) or $i_u=0$ (clusters denoted as $C^0$). 
Finally, we can initialize the algorithm as follows:
{
\begin{align*}
\alpha_G^c~=~
\begin{cases}
 \sigma \frac{|P| \cdot \textit{FPR}}{|C^0|} & \text{if}~c \in C^0 \land G=P  \\
 \sigma \frac{|P| \cdot (1-\textit{FPR})}{|C^0|} & \text{if}~c \in C^1 \land G=P  \\
 \sigma \frac{|N| \cdot (1-\textit{FNR})}{|C^1|} & \text{if}~c \in C^0 \land G=N  \\
 \sigma \frac{|N| \cdot \textit{FNR}}{|C^1|} & \text{if}~c \in C^1 \land G=N  \\
\end{cases} 
\end{align*}
}
Values of $\alpha_G^c$ encode 
beliefs 
of how many users from group $G$ should 
end up in cluster $c$ 
according to our trust in the  initial classification
based on NHST.
Parameter $\sigma$ 
balances between classification and clustering impact 
and informs 
how sure we are about the values of \textit{FPR} and \textit{FNR}.
For example, we use $\textit{FPR}=0.25$, $\textit{FNR}=0.4$ and $\sigma=1.0$.

The model fitting is performed in a standard way via EM, apart from two differences:
(1)~when calculating expectations new priors $\vec{\pi_G}$ are used, and
(2)~in the maximization step 
${\pi^c_G}$ are updated per group:
$$
\pi_G^c \sim \alpha_G^c + \sum_{u \in G} \gamma(z^c_{u})
$$
where $\gamma(z^c_{u})$ are posterior cluster responsibilities.

\xhdr{Prediction} 
Prediction of new users can be performed via co-clustering.
Specifically,
users for which we could not perform the statistical test
we include into the clustering as a new group ${X}$
with uninformative priors, for example $\alpha^c_X=1$.
The rest of the method remains unaltered. 

Co-clustering of users with badge and without badge
can improve classification results 
in both validation and prediction, but
the data distributions must be similar in terms of 
groupping attracted and not-attracted users.
This assumption may be hard to ensure for real data.
Therefore, to improve robustness and prediction quality,
we propose to first cluster users with badge using the above procedure and then employ clustering results to train a standard classifier with better generalization properties, for example logistic regression. 

\section{Synthetic Data Evaluation}
\label{sec:synthetic}

In this section, we compare with the help of synthetically generated
data the effectiveness of the proposed approaches for validation and
prediction of users attracted by the badge.
\begin{figure*}[ht]
\centering             
\captionsetup[subfigure]{labelformat=empty}
\captionsetup[subfigure]{justification=centering}

\subfloat[]{
\hspace*{-0.4cm}
\textbf{Poisson clustering}
}\hspace*{1.8cm}
\subfloat[]{
\textbf{NHST bootstrap}
}\hspace*{1.8cm}
\subfloat[]{
\textbf{2-phase bootstrap}
}\hspace*{0.0cm}

\vspace{-0.5cm}

\rotatebox{90}{ \hspace*{0.5cm} \textbf{{Validation}}}
\subfloat[]{
\includegraphics[width=0.3\textwidth]{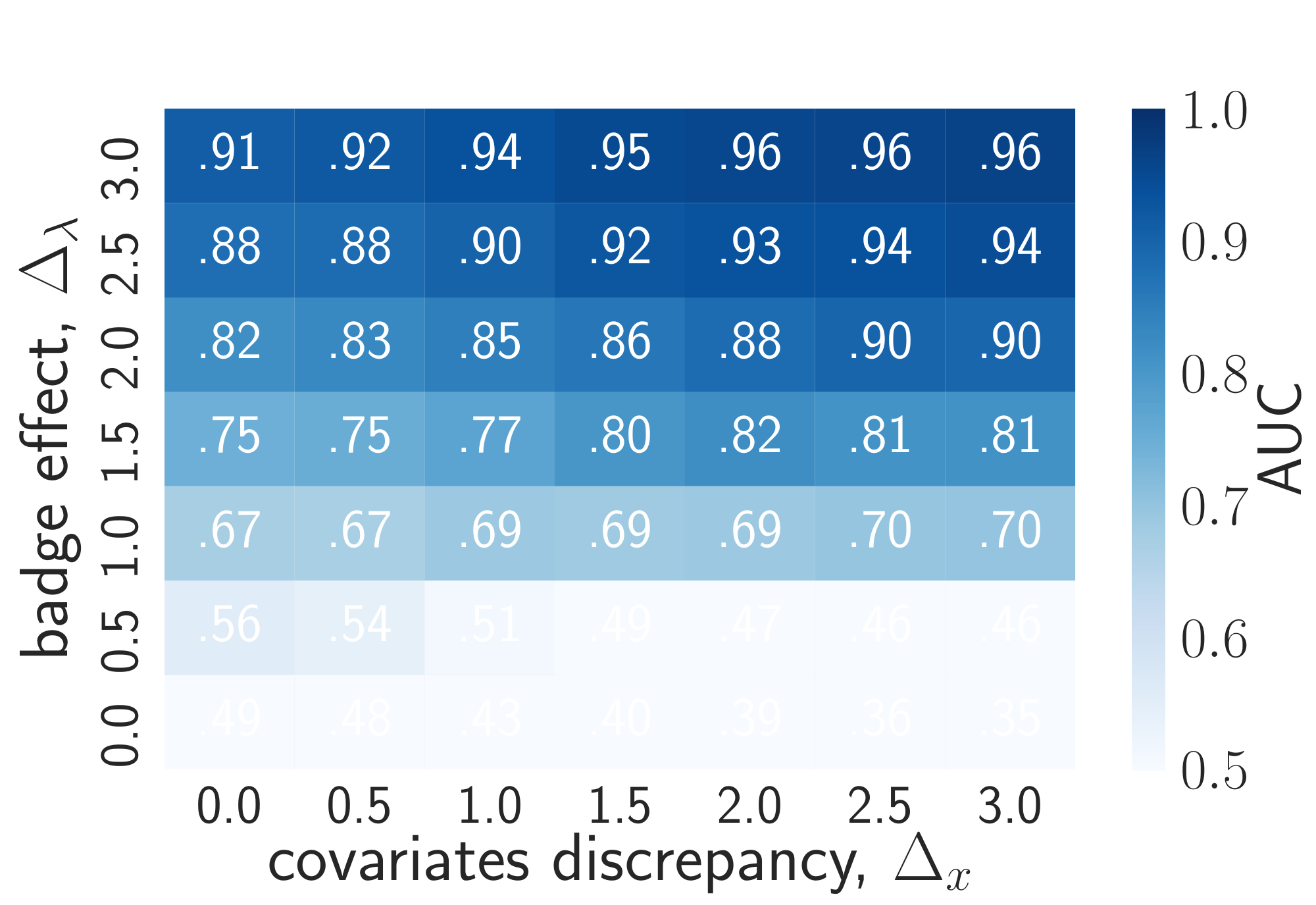} 
}
\subfloat[]{
\includegraphics[width=0.3\textwidth]{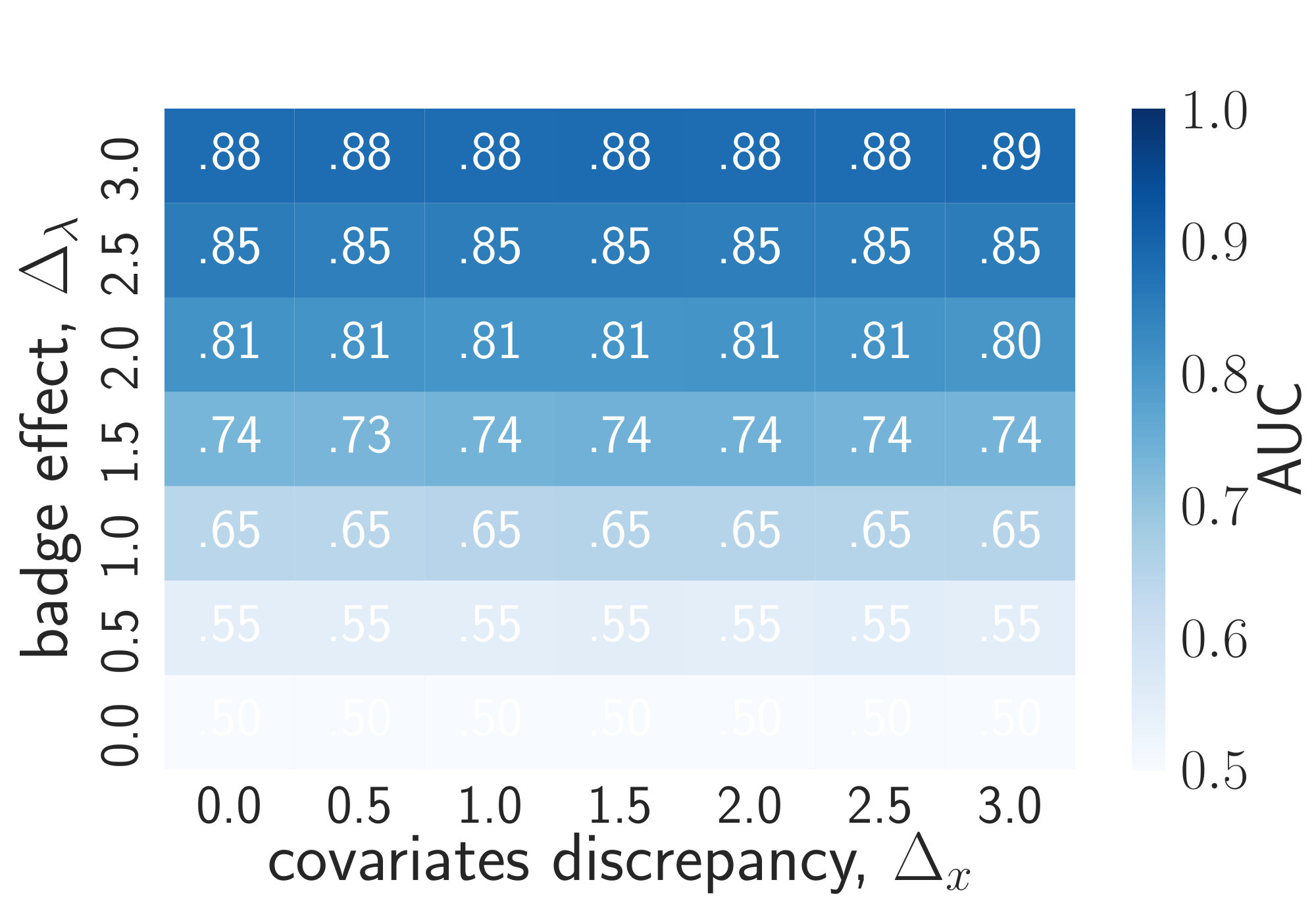}
}
\subfloat[]{
\includegraphics[width=0.3\textwidth]{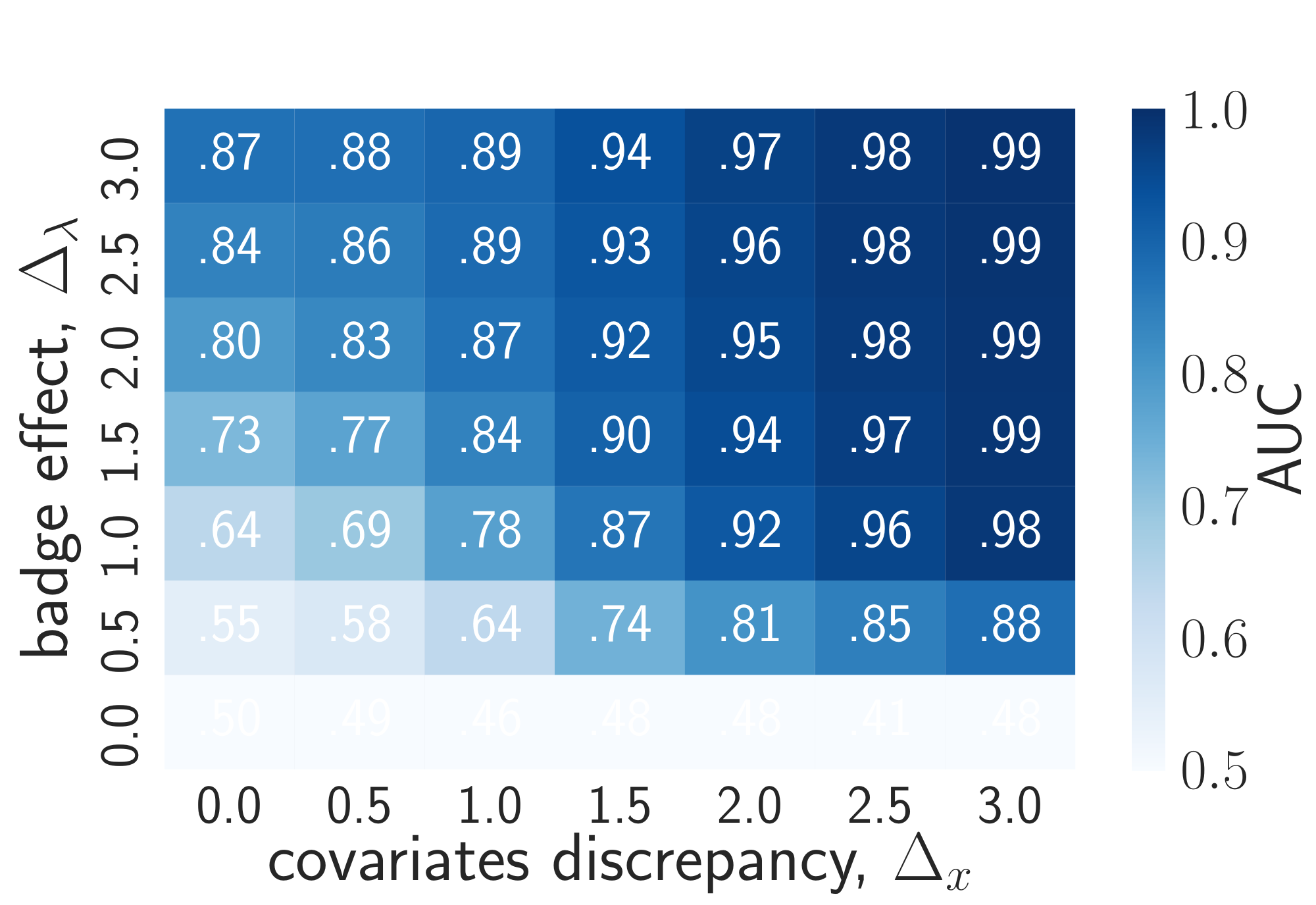}
}

\vspace{-0.5cm}

\rotatebox{90}{ \hspace*{0.5cm} \textbf{{Prediction}}}
\subfloat[]{
\includegraphics[width=0.3\textwidth]{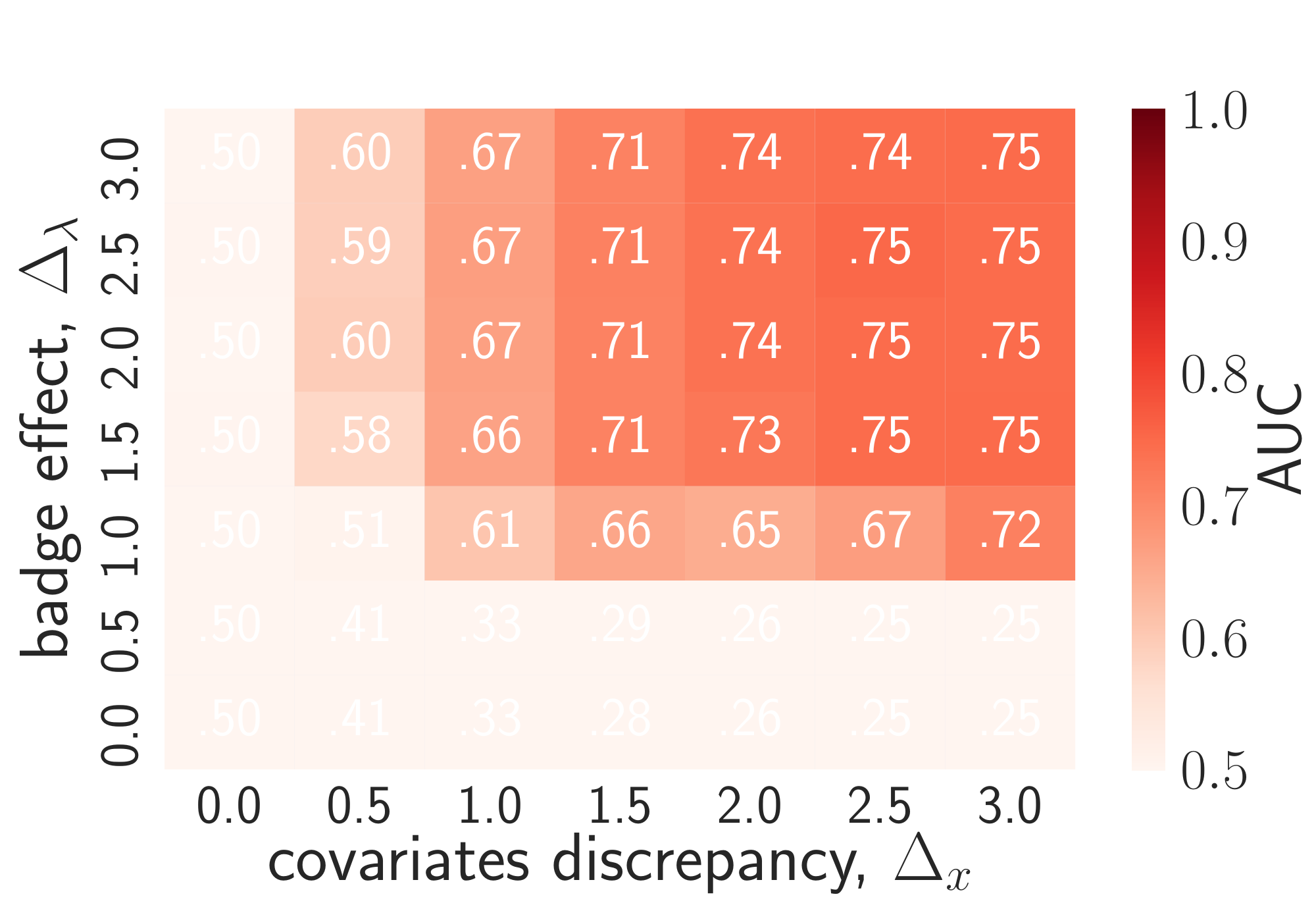} 
}
\subfloat[]{
\hspace{5cm}
}
\subfloat[]{
\includegraphics[width=0.3\textwidth]{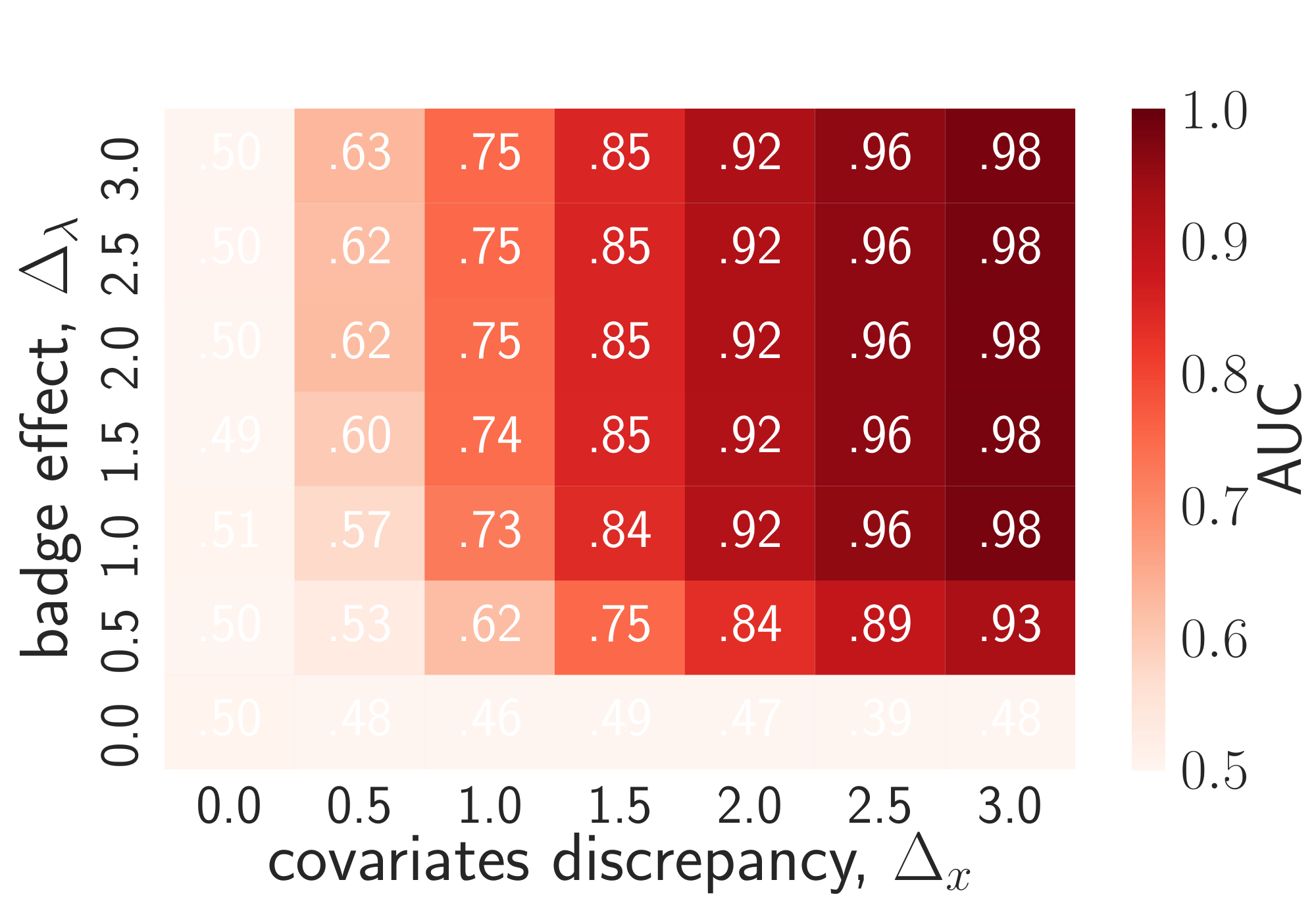}
}

\caption{Performance (average AUC) of our methods on synthetic data 
against badge effect ($\Delta_\lambda$) and covariates strength ($\Delta_x$). 
The top row shows the validation of badges' causal effect on users with badge (\ie, having sufficient $\{t_u\}$). The bottom row shows the performance for new users (\ie, with only  $\vec{x_u}$ employed).
}
\label{fig:synthetic_results}
\end{figure*}

\xhdr{Basic Setting}
We simulate the behavior of $N=1000$ users:
$N/2$ users with both temporal dynamics $\{t_u\}$ and covariates $\vec{x_u}$, and $N/2$ users with only covariates $\vec{x_u}$, 
that imitate new users.
For each user we assign a latent variable $i_u$: with probability $\pi$: $i_u=1$ and with probability $1-\pi$: $i_u=0$.

The users' temporal traces $\{t_u\}$ we sample according to
intensities expressed in Eq.~\ref{eq:basic-model1}. 
The intensities $\lambda^0(u)$ and $\lambda_1^1(u)$ 
we draw according to Eq.~\ref{eq:robust-survival-model} 
where we fix variances $\textit{Var}(\lambda^0(u))~=~\textit{Var}(\lambda_1^1(u))~=~25$
and means $E(\lambda^0(u))=10$, $E(\lambda_1^1(u))=10-\Delta_\lambda$,
and
intensity $\lambda_0^1(u)$ we fix respectively to $\lambda_0^1(u) = \lambda_1^1(u)+2\Delta_\lambda$. 
The parameter
\textit{$\Delta_\lambda$ controls 
the strength of the simulated badge effect}.
In the basic setting,
randomness in individual user
temporal trace $\{t_u\}$
appears due to point processes sampling procedure.

Users are independent, and therefore 
without loss of generality 
we can assume start times for all users $s_u=0$. 
Furthermore,
we set badge awarding time to $b_u=100/\lambda^0(u)$ for users with $i_u=0$,
and $b_u=100/\lambda_0^1(u)$ for  users with $i_u=1$. 
User end times we set to $e_u = b_u+u\cdot b_u$ ($u\sim~U[0,1]$).

We sample user features from bivariate (=two features per user) normal distributions:
\begin{align*}
\vec{x_u} ~\sim~
{
\begin{cases}
N(0, \Sigma) & \text{when}~ i_u=0 \\
N(\Delta_x \cdot s_{max} \vec{v}_{max}, \Sigma) &  \text{otherwise}
\end{cases}
}
\end{align*}
where covariance matrix 
{
$\Sigma~\sim~\text{Wishart}~(10,~\begin{bmatrix}
   2 & 1\\
   1 & 2\\
\end{bmatrix}
)$},
$s_{max}$ is the largest singular value of $\Sigma$ 
corresponding to eigenvector $\vec{v}_{max}$,
and 
\textit{$\Delta_x$ controls discrepancy between features of users from different groups}.

\xhdr{Disturbed Data Setting} We study
the robustness 
of our methods 
by simulating
disturbed data,
\eg, 
temporal fluctuations
in user intensities.
In particular, 
we add
(typical for real data)
temporal trend, \ie,
$\lambda^{i_u}_u(t)$ 
becomes
$\lambda^{i_u}_u(t)(1+A t)$,
where $A$ controls the extent of the simulated fluctuation.

\begin{figure}[t]
\centering
\captionsetup[subfigure]{labelformat=empty}
\captionsetup[subfigure]{justification=centering}

\subfloat[\textbf{Validation}]{
\hspace{-0.45cm}
\includegraphics[width=0.35\textwidth]{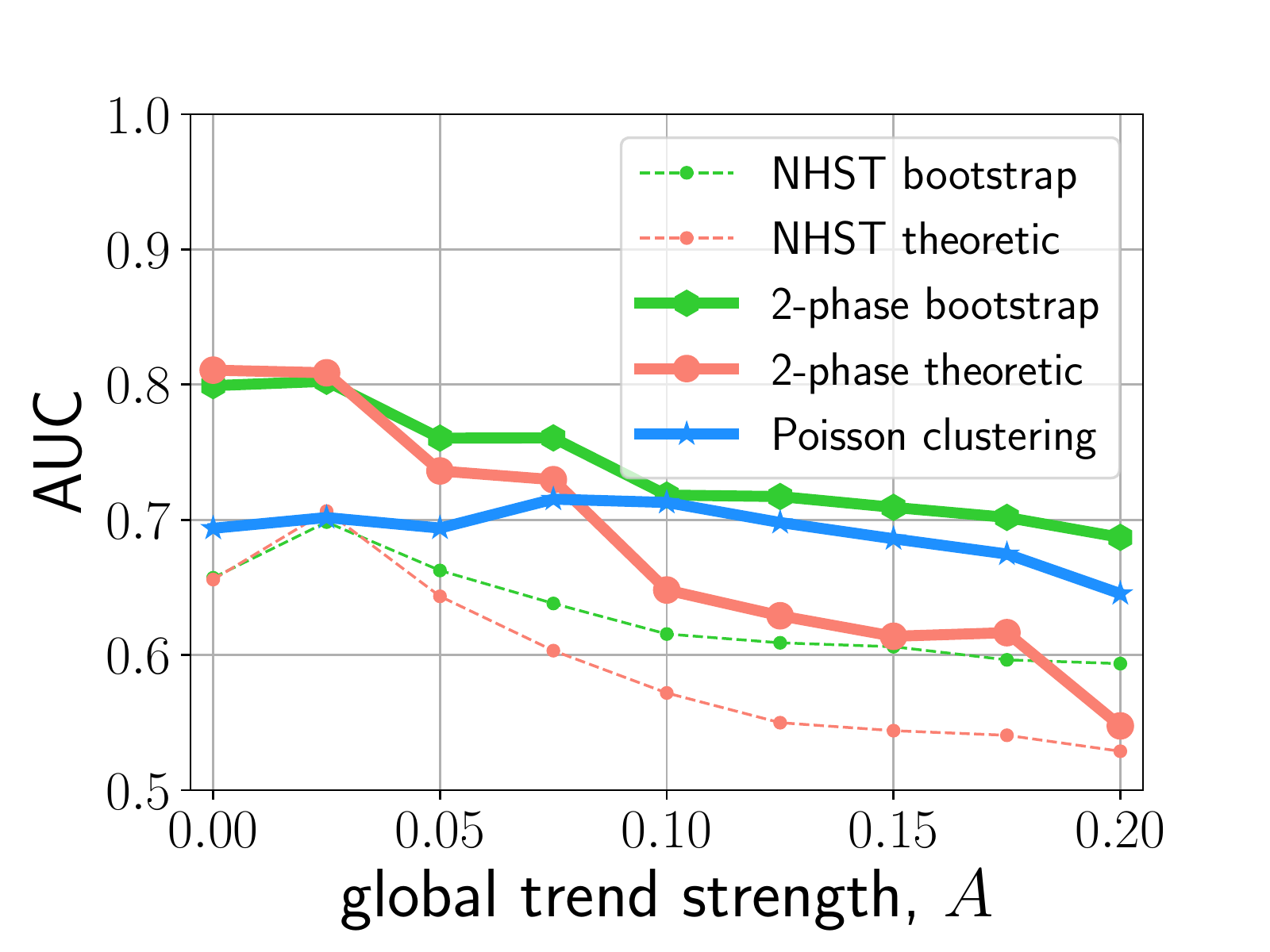}
}
\subfloat[\textbf{Prediction}]{
\hspace{-0.705cm}
\includegraphics[width=0.35\textwidth]{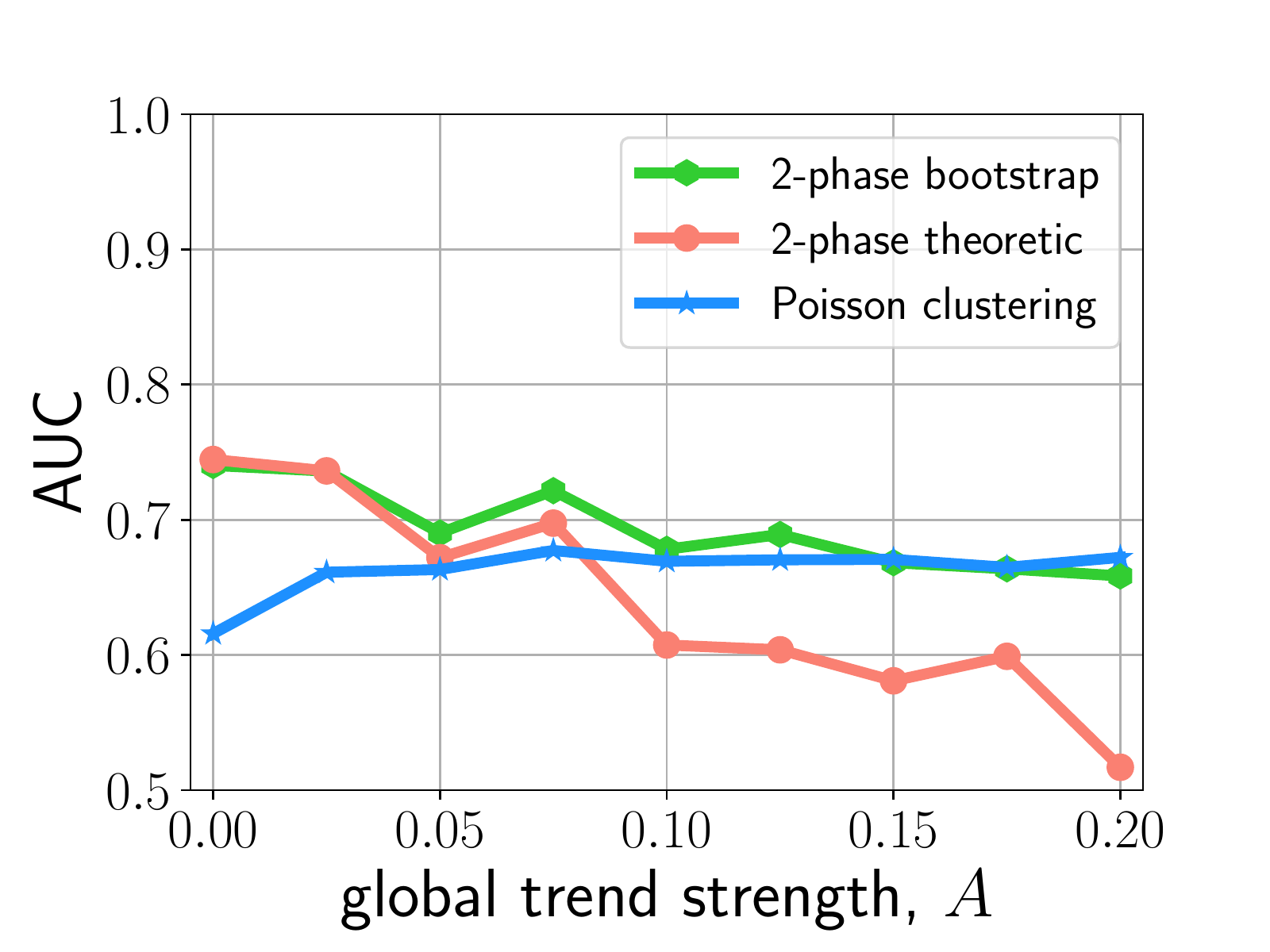}
}
\caption{Robustness 
against temporal fluctuations (\eg,~global linear trend; $\Delta_\lambda=\Delta_x=1$, $\pi=0.5$). }
\label{fig:synthetic_trend}

\end{figure}
\begin{figure}[t]
\centering
\captionsetup[subfigure]{labelformat=empty}
\captionsetup[subfigure]{justification=centering}

\subfloat[\textbf{Validation}]{
\hspace{-0.45cm}
\includegraphics[width=0.35\textwidth]{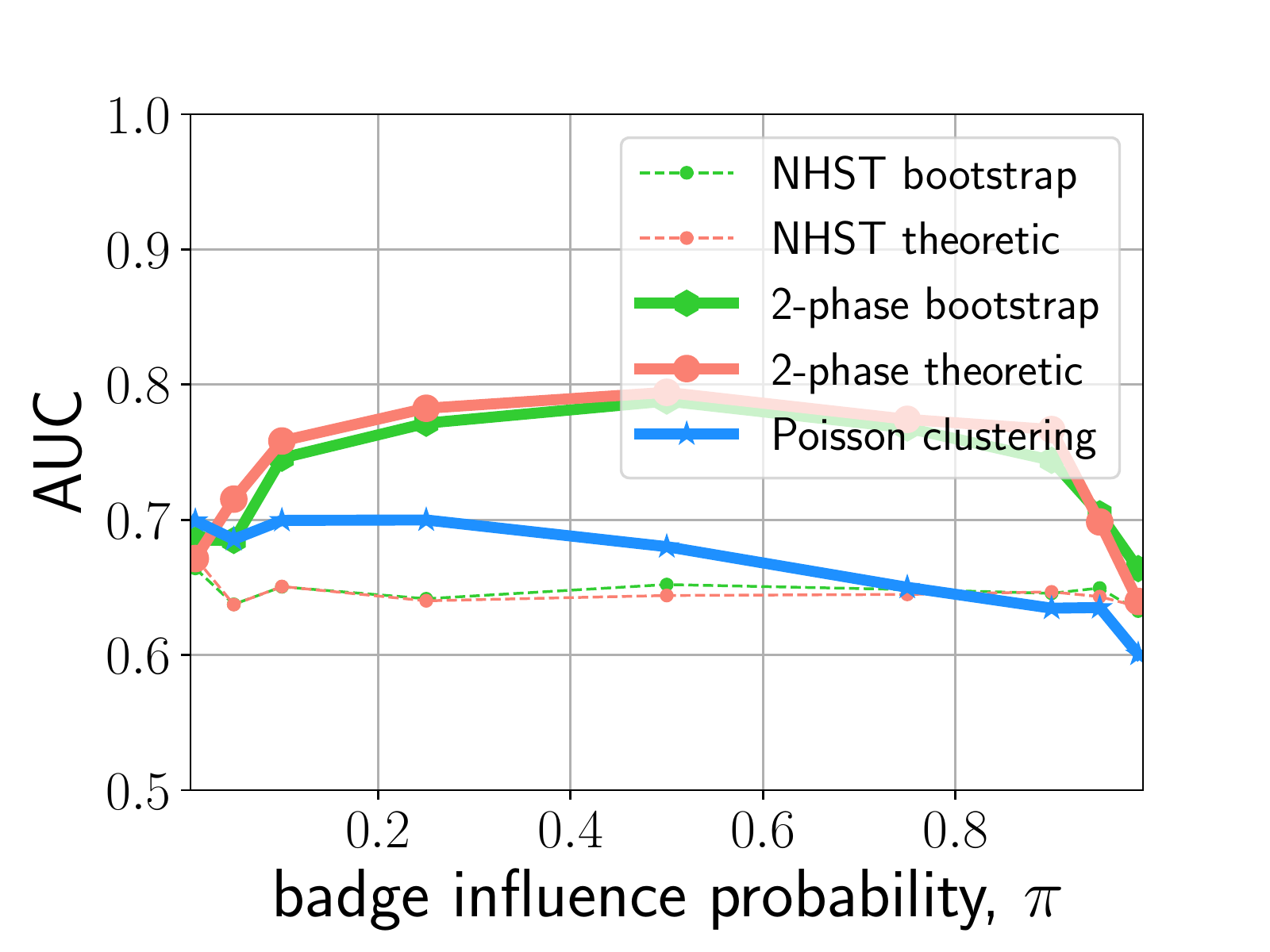}
}
\subfloat[\textbf{Prediction}]{
\hspace{-0.705cm}
\includegraphics[width=0.35\textwidth]{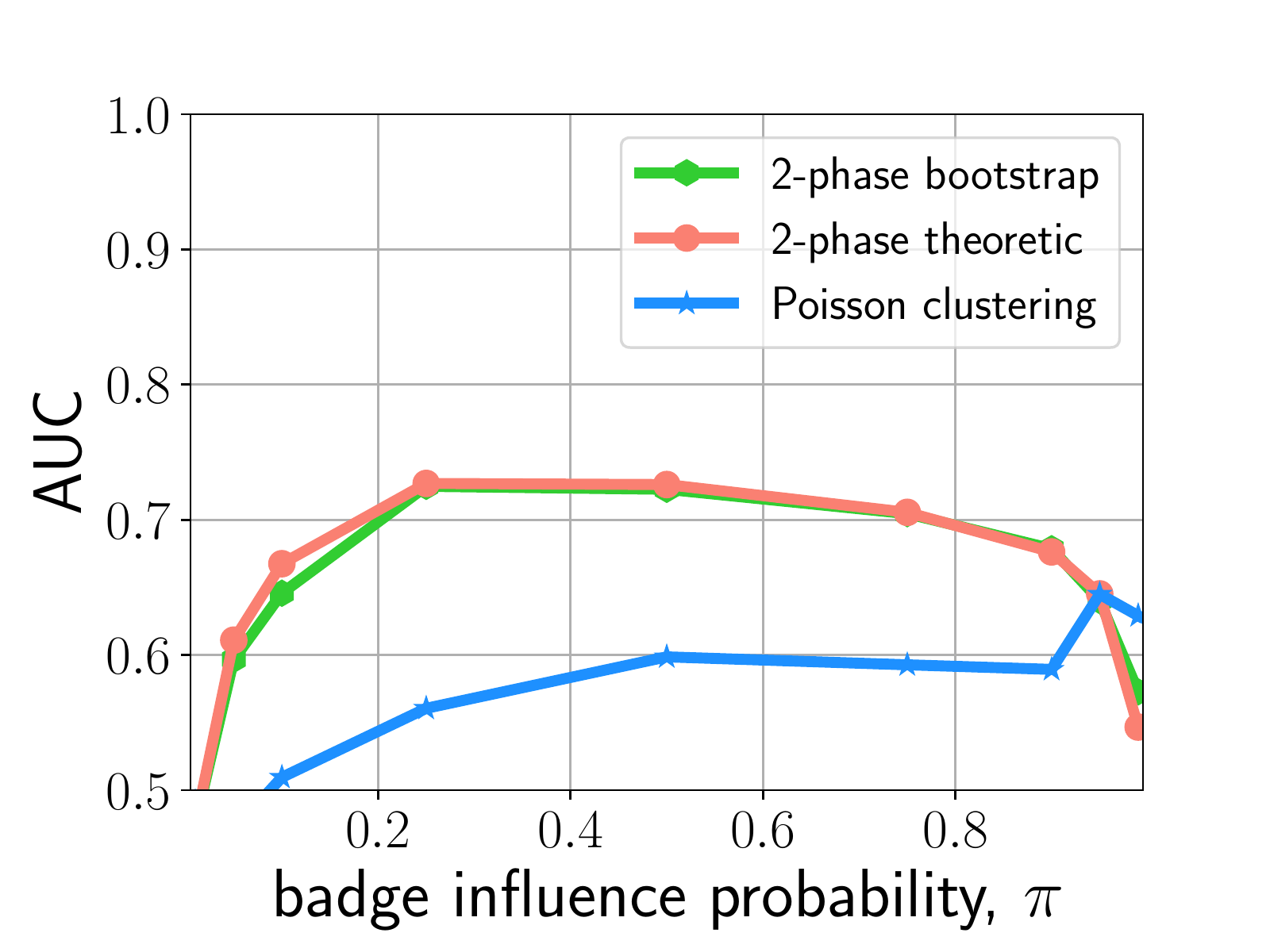}
}

\caption{Sensitivity to class imbalance ($\Delta_\lambda=\Delta_x=1$).}
\label{fig:synthetic_imbalance}
\end{figure}

\xhdr{Evaluation} 
In contrast to what is the case for real data,
for synthetic data 
we know user attitude towards the badge ($i_u$) 
and therefore
we can evaluate
prediction results 
against it. 
Specifically,
we employ \textit{Area Under Curve} (AUC) 
that accounts for uncertainty 
in our methods predictions.
We measure AUC separately for users with full information (user
validation problem) and for users with limited data (user prediction problem).
Every experiment we repeat $20$ times and then average results.

\xhdr{Results}
Figure~\ref{fig:synthetic_results} summarizes the simulation results 
in the basic setting
for 
varying 
badge effects ($\Delta_\lambda$) and 
clusterization levels ($\Delta_\lambda$).
The methods based on 
our
 \textit{2-phase} procedure (we show results only for \textit{bootstrap} variant; results for \textit{theoretic} variant in the basic setting are identical)
  improve over the basic \textit{NHST} classification
 and
  have a superior performance over
 \textit{Poisson (processes) clustering}.
Poisson clustering
fails
when intensity differences between user classes are small (\eg,~$\Delta_\lambda\leq0.5$).
We found  that
the method
in the 
considered
variant (with logistic regression in Eq.~\ref{eq:functional})
does not 
benefit sufficiently
from differences between users' covariates ($\Delta_x$)
and
as a result 
underestimates the 
probabilities of badges having effect.

Fluctuations in the temporal data lead to
performance degradation.
For example, 
Figure~\ref{fig:synthetic_trend} shows how AUC is decreasing 
when divergence from the models (controlled by trend $A$) is increasing.
The most affected method is 
\textit{2-phase theoretic}
(2-phase procedure with theoretic estimation of the test statistic distribution).
The intermediate results 
(\textit{NHST theoretic})
confirm that the method tend to overestimate badge influence 
when fluctuations are strong.
However
the least affected
is 
\textit{Poisson processes clustering},
it never outperforms \textit{2-phase bootstrap} - the robust variant of the our 2-phase procedure.

Finally, we investigate our methods in terms of sensitivity to class imbalance (Figure~\ref{fig:synthetic_imbalance}). 
Class imbalance has a low impact on validation performance.
It happens because
the main indicator
of the badge influence on user
is a change in individual user dynamics around the badge awarding time
and covariates are only `regularizers'.
On the other hand, the prediction relies entirely on covariates
and
if one class is 
underrepresented
covariates distributions are poorly fitted and prediction fails, \eg, AUC approaches $0.5$ for both very small and large $\pi$.

\section{Real Data Experiments}
\label{sec:real}

In this section, we investigate 
the effectiveness 
of
our methods 
when applied to real data, \ie,
two sample badges from 
a popular
Q\&A platform.

\xhdr{Data Description and Preprocessing}
In the real-data experiments we used a Stack Overflow dataset~\footnote{https://archive.org/details/stackexchange}, 
that contains 
timestamped events 
from
between July 2008 and September 2014
and some basic information about users.
In particular, 
we used the following user features and statistics as badge \textit{covariates}:
\begin{itemize}[noitemsep,nolistsep,leftmargin=0.7cm]
\item[--] user age and location 
\item[--] total number of user page views, posted comments, upvotes and downvotes
\end{itemize}
From location we extracted 
city and state names that 
we independently embedded using a pre-trained word2vec model\footnote{https://code.google.com/archive/p/word2vec/}. 
Embeddings were clustered separately into 5+5 clusters
using k-means and distances to cluster centers were subsequently used as covariates: 5 for city and 5 for state. 
We transformed user statistics by applying the following 
transformation: $x \rightarrow \log(x+1)$. 
We also filtered out users with incomplete records.

We demonstrate the effectiveness of the proposed approaches for
two sample \textit{threshold badges}\footnote{{https://meta.stackexchange.com/questions/67397/}}: 
\begin{itemize}[noitemsep,nolistsep,leftmargin=0.7cm]
\item[--] \badge{Research Assistant}: awarded to users who edited at least $50$ wiki sites describing tags (wiki tag edits). Users with reputation\footnote{https://stackoverflow.com/help/whats-reputation} $1500$ or higher can perform these actions.
\item[--] \badge{Copy Editor}: awarded to users who performed a total of $500$ post (\eg, question or answer) edits. Users with reputation $100$ or higher can perform these actions.
\end{itemize}

\begin{figure*}[t!]
\centering             
\captionsetup[subfigure]{labelformat=empty}
\captionsetup[subfigure]{justification=centering}

\subfloat[]{
\textbf{Poisson clustering}
}\hspace*{2.05cm}
\subfloat[]{
\textbf{NHST bootstrap}
}\hspace*{2.05cm}
\subfloat[]{
\textbf{2-phase bootstrap}
}\hspace*{0.25cm}

\vspace*{-0.5cm}

\rotatebox{90}{ \hspace*{0.25cm} {\small\badge{{Research Assistant}}}}
\subfloat[]{
\includegraphics[width=0.31\textwidth]{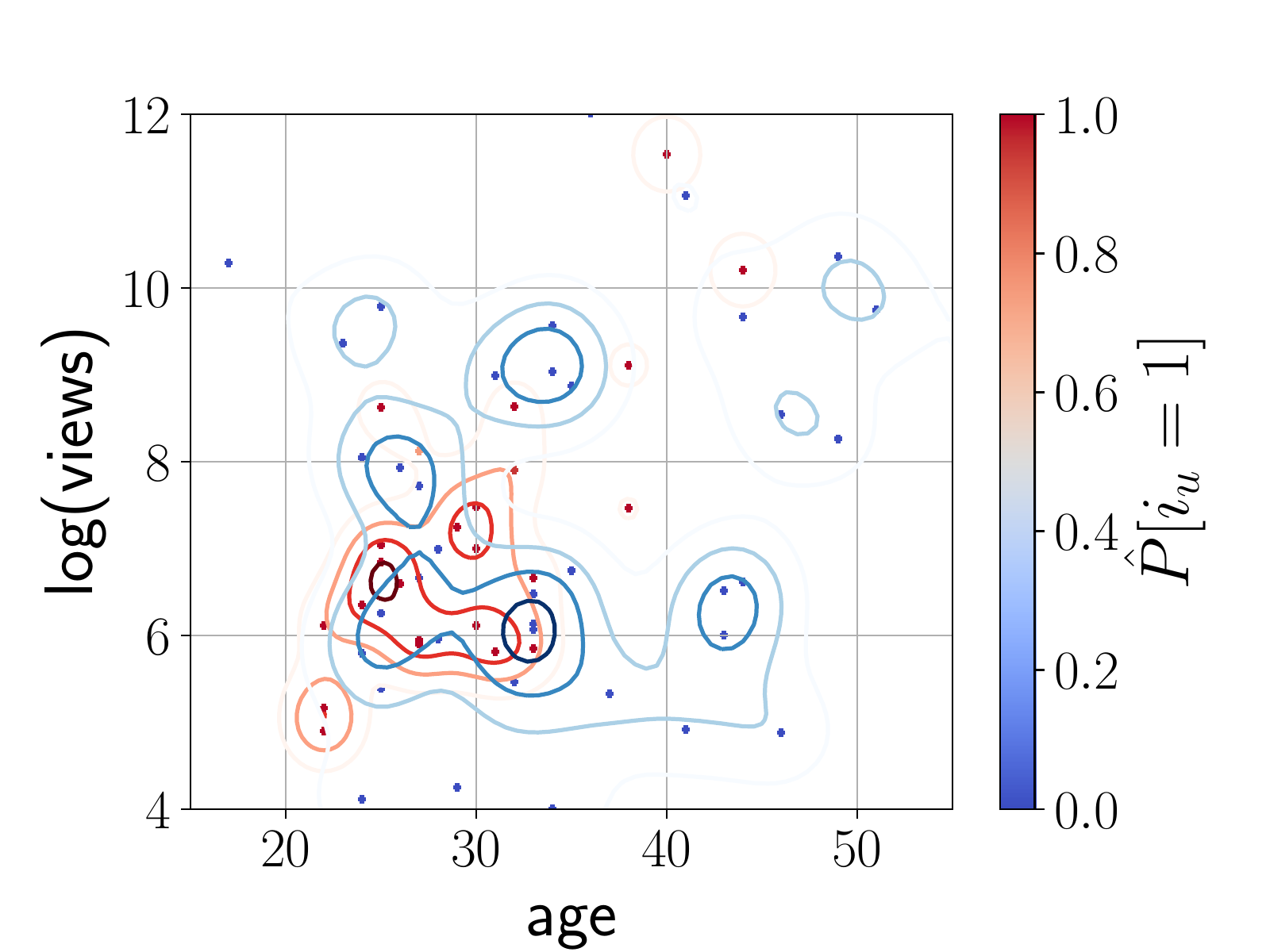} 
}
\subfloat[]{
\includegraphics[width=0.31\textwidth]{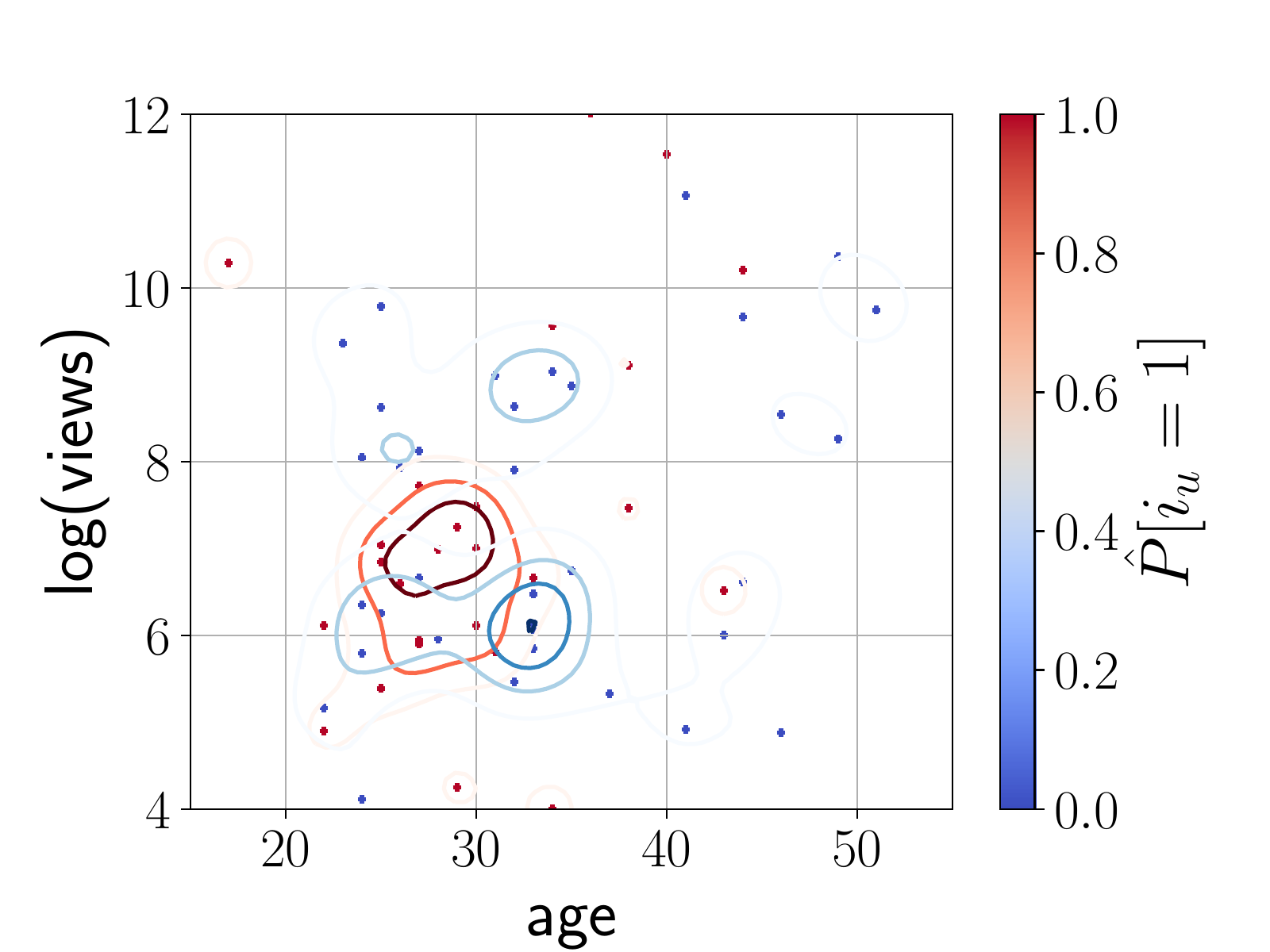}
}
\subfloat[]{
\includegraphics[width=0.31\textwidth]{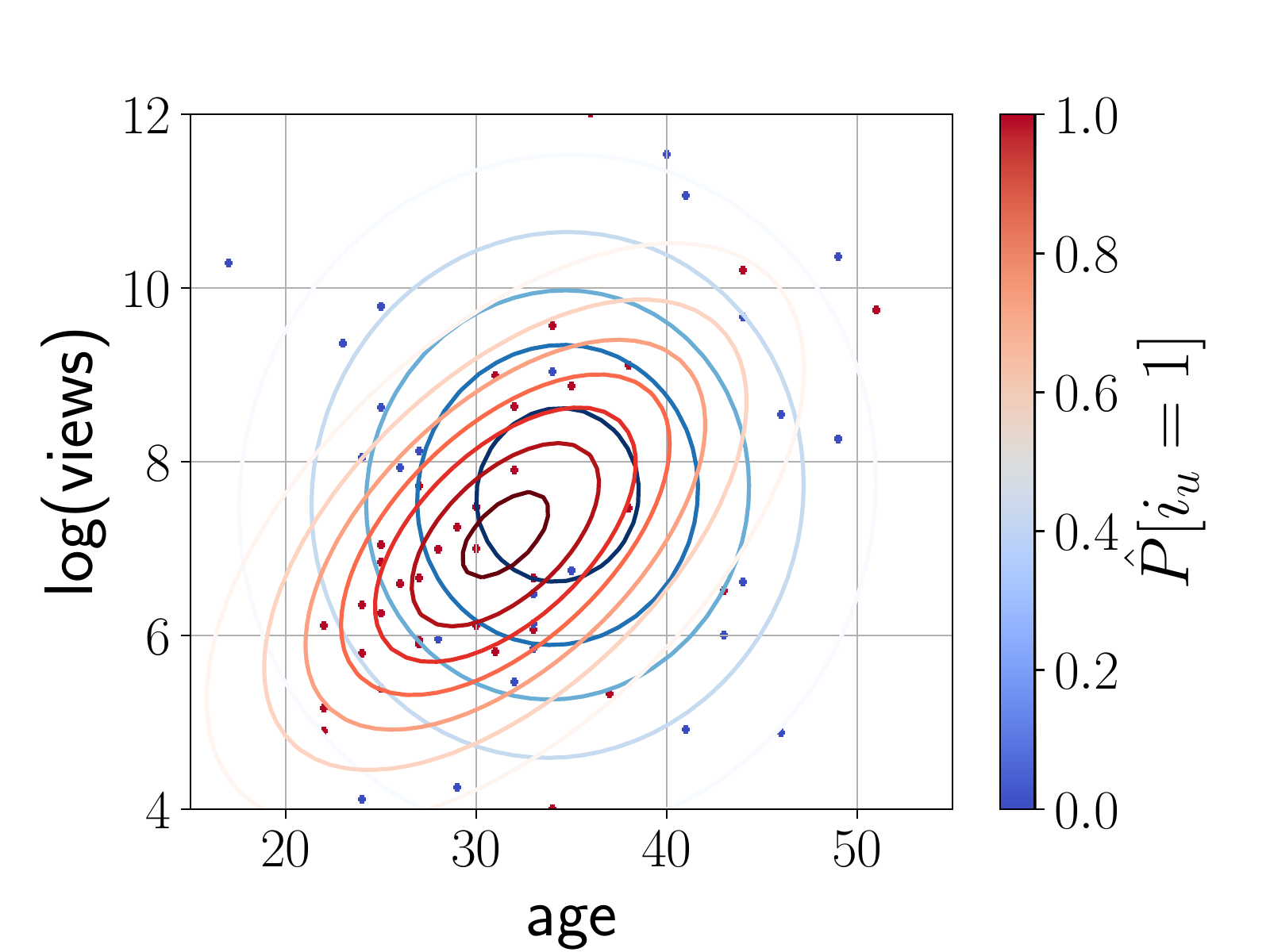}
}

\vspace*{-0.5cm}

\rotatebox{90}{ \hspace*{0.75cm} {\small\badge{{Copy Editor}}}}
\subfloat[]{
\includegraphics[width=0.31\textwidth]{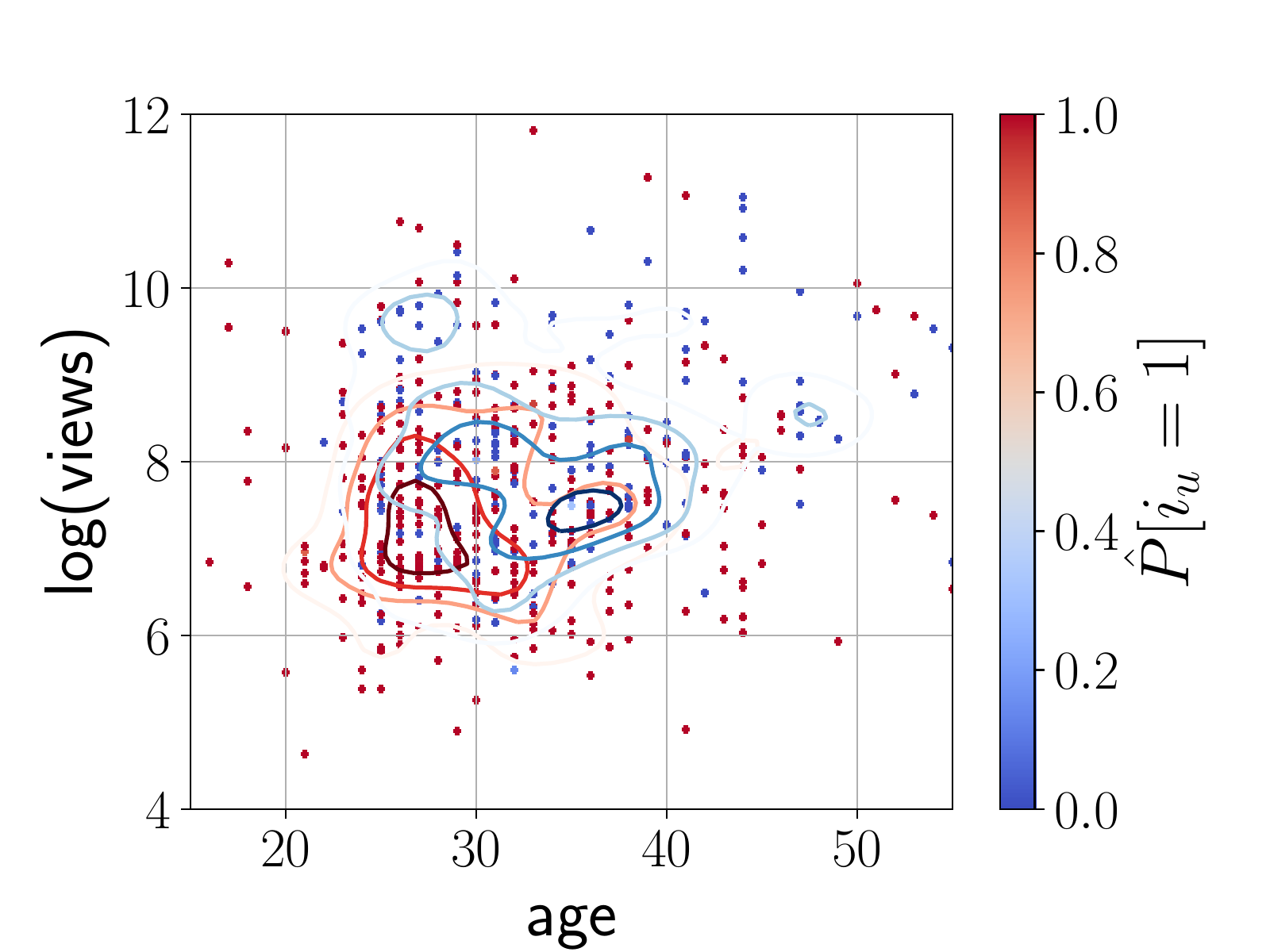} 
}
\subfloat[]{
\includegraphics[width=0.31\textwidth]{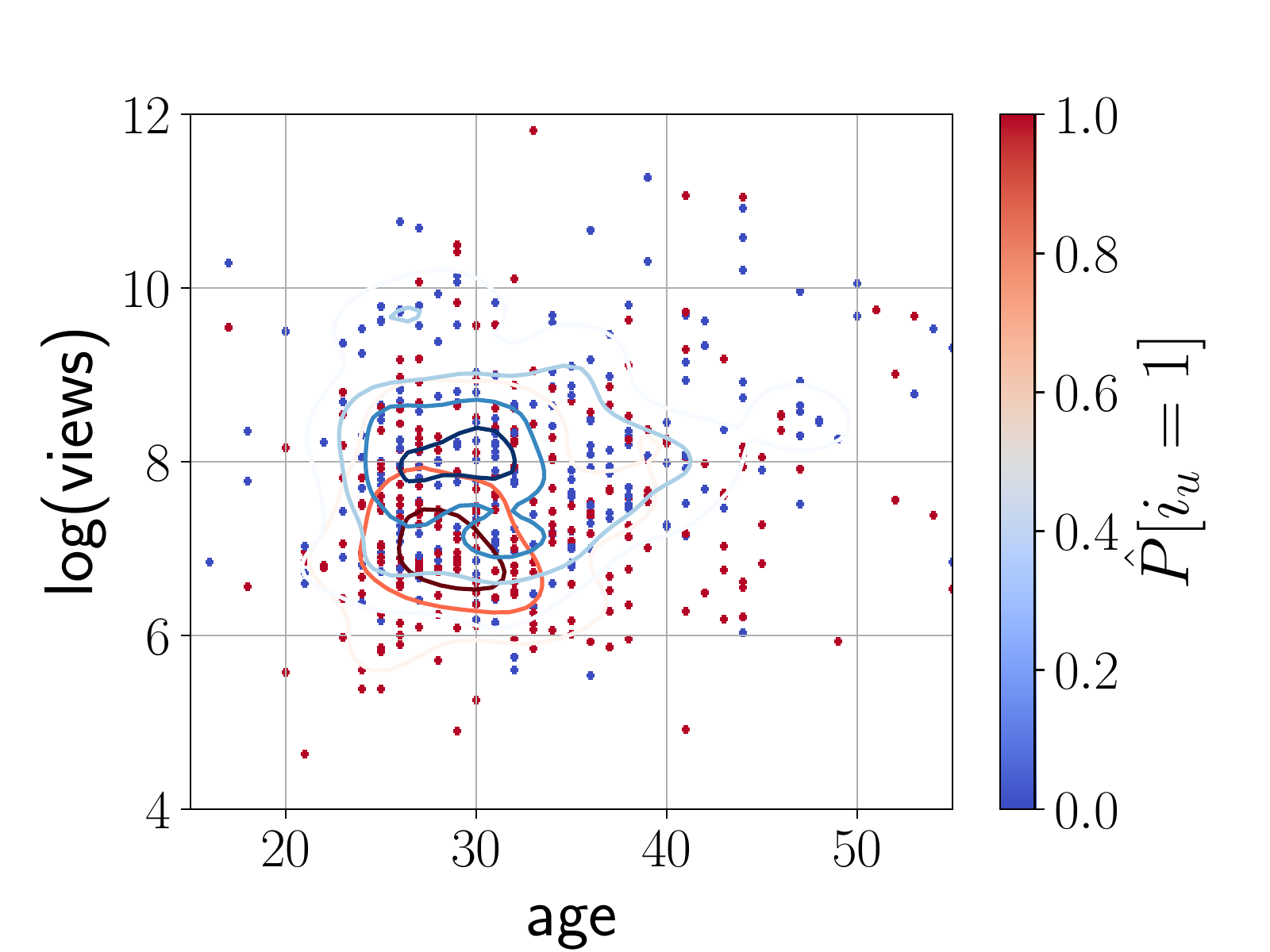}
}
\subfloat[]{
\includegraphics[width=0.31\textwidth]{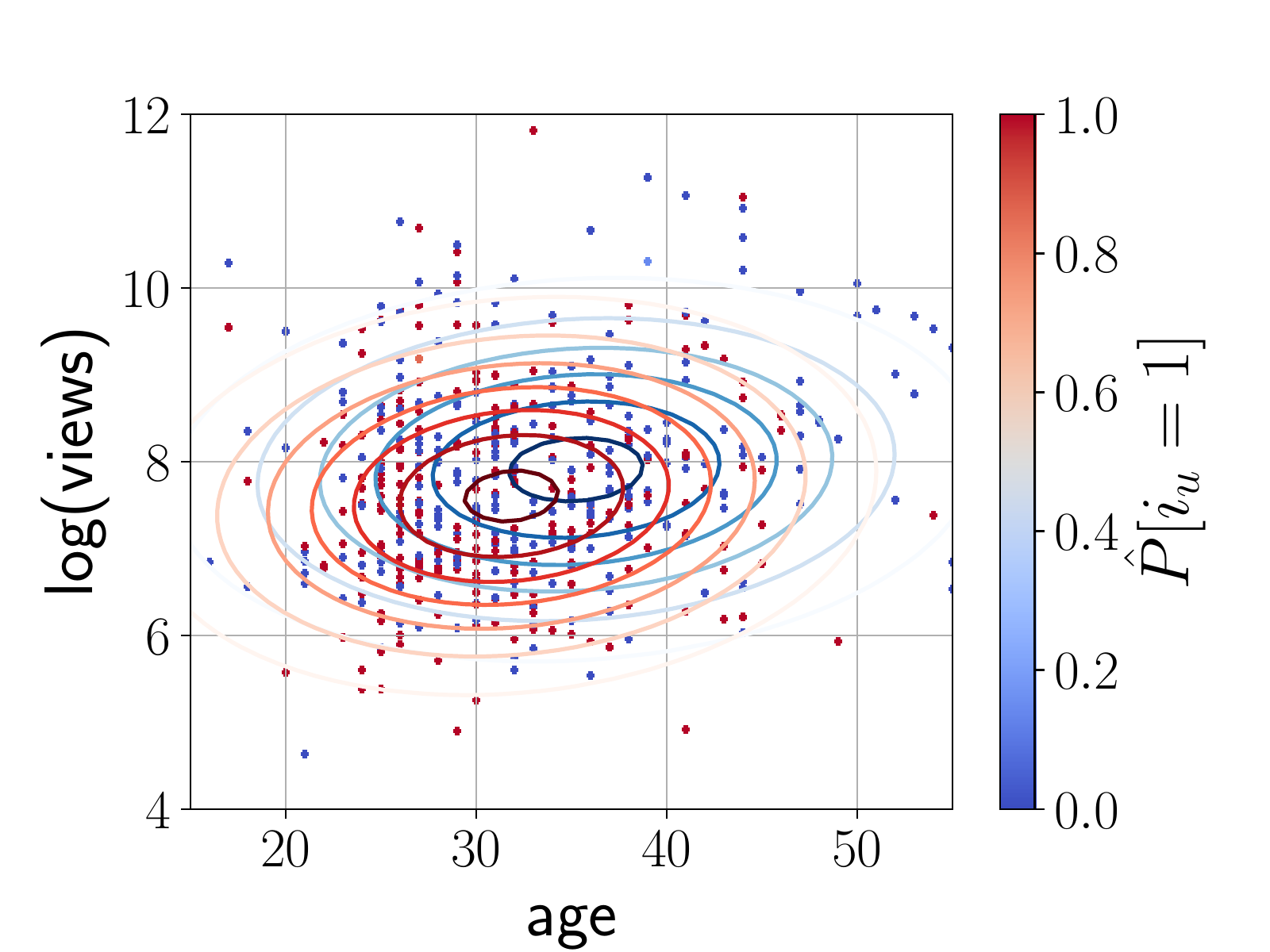}
}

\caption{Validation of the effect of two badges: \badge{{Research
      Assistant}} (top) and \badge{{Copy Editor}} (bottom). Users are
  projected onto a two-dimensional space of \textit{age} and \textit{log number of views}, with badge effect represented with color (\texttt{red}=influenced, \texttt{blue}=badge awarded by chance).
}
\label{fig:ral_results}
\end{figure*}

\xhdr{Results}
Figure~\ref{fig:ral_results} illustrates validation results from Poisson processes clustering and 2-phase bootstrap alongside with intermediate results from NHST bootstrap. 
The classification results from different methods agree (=badge effect probability either larger than $0.5$ or smaller than $0.5$ in both cases) to a high degree. 
For example, for \badge{Research Assistant} 
we observe 70\% agreement  between Poisson clustering and 2-phase bootstrap. With $\text{p-value}<0.001$ we can reject the hypothesis that it happens by chance. Similarly, for \badge{Copy Editor} we report $\text{p-value}=0.016$. 

The validation results
 suggest that only about half (\ie, 58\% for \badge{Research Assistant} and 47\% for \badge{Copy Editor} according to 2-phase bootstrap) of the users intentionally 
 performed actions needed to receive the badge. 
Prediction results are less conclusive. 
However, in 2-phase bootstrap classification for \badge{Research Assistant}  and for \badge{Copy Editor}
we got respectively 53\% and 39\% of users potentially attracted to the badge (only users with sufficient reputation included),
Poisson processes clustering classified all new users as unlikely interested in the badges. 
We presume that this can happen due to differences between data distributions  of users with and without badge
that are handled differently by the methods.

Examination of fitted models can give deeper insights into how user characteristics relate to badges influence. 
In particular, 
we ranked covariates according to \textit{Kullback–Leibler divergence} between both user classes ($i_u=0/1$) 
and reported means of the respective distributions.  
We observed that features derived from location best discriminate
between classes. For example, we discovered that users located in
the USA 
were getting badges more often by chance -- due to their natural high activeness.
On the other hand, users from East Europe and India were more mercenary -- 
their behavior was more often driven by the perspective of a badge reward. 
Similarly, 
we found out that
younger users 
on average
were more goal-oriented than older ones.

\section{Conclusions}
\label{sec:discussion}

Badges are a 
popular motivational mechanism used in social media sites.
However, due to complexity of these environments,
the belief that they really work, 
\ie, are incentivizing users to perform certain actions,
is hard to verify
and
until recently there were no tools for that.
To address this problem 
we designed and evaluated 
two approaches
to verify individual users attraction towards badges. 
The proposed methods applied 
to
real data from
Stack Overflow
let us 
to gain interesting 
insights 
about users who earn badges.
In particular,
in contradiction to previous beliefs
we discovered 
that
many of them 
receive badges by chance, 
having no prior intention of it.

Our work
can be extended in
many ways. 
For example,
it would be interesting to see 
how more advanced features,
like temporal features covering user evolution on early stage,
can improve
the performance of our methods.
Furthermore, we focused our research on threshold badges 
(that are the most popular ones)
but
there are many other interesting designs 
(for example badges associated to limited resources)
for which the problem of influence validation remains open.
Finally, 
we believe that 
our results should affect 
how 
badges are designed 
and help in making them more effective.
\section*{Acknowledgements} 
We thank 
Manuel Gomez-Rodriguez for inspiring us 
to perform this research,
Eliezer de Souza da Silva
for useful discussions,
and both of them along with Sean Chester
for critical review of the publication.

{
\bibliographystyle{abbrv}
\bibliography{refs}

\begin{thebibliography}{10}

\bibitem{abramovich2013badges}
S.~Abramovich, C.~Schunn, and R.~M. Higashi.
\newblock Are badges useful in education?: It depends upon the type of badge
  and expertise of learner.
\newblock {\em Educational Technology Research and Development},
  61(2):217--232, 2013.

\bibitem{anderson2013steering}
A.~Anderson, D.~Huttenlocher, J.~Kleinberg, and J.~Leskovec.
\newblock {Steering user behavior with badges}.
\newblock In {\em Proceedings of the 22nd international conference on World
  Wide Web}, 2013.

\bibitem{aral2012identifying}
S.~Aral and D.~Walker.
\newblock Identifying influential and susceptible members of social networks.
\newblock {\em Science}, 337(6092):337--341, 2012.

\bibitem{bishop}
C.~M. Bishop.
\newblock {\em Pattern Recognition and Machine Learning (Information Science
  and Statistics)}.
\newblock Springer-Verlag New York, Inc., Secaucus, NJ, USA, 2006.

\bibitem{FM7299}
B.~Bornfeld and S.~Rafaeli.
\newblock Gamifying with badges: A big data natural experiment on stack
  exchange.
\newblock {\em First Monday}, 22(6), 2017.

\bibitem{Colquhoun140216}
D.~Colquhoun.
\newblock An investigation of the false discovery rate and the
  misinterpretation of p-values.
\newblock {\em Royal Society Open Science}, 1(3), 2014.

\bibitem{daley2002introduction}
D.~Daley and D.~Vere-Jones.
\newblock {\em An Introduction to the Theory of Point Processes: Volume I:
  Elementary Theory and Methods}.
\newblock Probability and Its Applications. Springer, 2002.

\bibitem{easley2016incentives}
D.~Easley and A.~Ghosh.
\newblock Incentives, gamification, and game theory: an economic approach to
  badge design.
\newblock {\em ACM Transactions on Economics and Computation}, 4(3):16, 2016.

\bibitem{ghosh2011incentivizing}
A.~Ghosh and P.~McAfee.
\newblock Incentivizing high-quality user-generated content.
\newblock In {\em Proceedings of the 20th international conference on World
  wide web}, pages 137--146, 2011.

\bibitem{gibson2015digital}
D.~Gibson, N.~Ostashewski, K.~Flintoff, S.~Grant, and E.~Knight.
\newblock Digital badges in education.
\newblock {\em Education and Information Technologies}, 20(2):403--410, 2015.

\bibitem{HAMARI2017469}
J.~Hamari.
\newblock Do badges increase user activity? a field experiment on the effects
  of gamification.
\newblock {\em Computers in Human Behavior}, 71:469 -- 478, 2017.

\bibitem{hamari2014does}
J.~Hamari, J.~Koivisto, and H.~Sarsa.
\newblock Does gamification work?--a literature review of empirical studies on
  gamification.
\newblock In {\em System Sciences (HICSS), 2014 47th Hawaii International
  Conference on}, pages 3025--3034. IEEE, 2014.

\bibitem{hogg1995introduction}
R.~V. Hogg and A.~T. Craig.
\newblock {\em Introduction to mathematical statistics}.
\newblock Prentice Hall, 1995.

\bibitem{ImmStoSyr15}
N.~Immorlica, G.~Stoddard, and V.~Syrgkanis.
\newblock Social status and badge design.
\newblock In {\em Proceedings of the 24th international conference on World
  Wide Web}, 2015.

\bibitem{1707.08160}
T.~Kusmierczyk and M.~Gomez-Rodriguez.
\newblock Harnessing natural experiments to quantify the causal effect of
  badges. \textit{arXiv:1707.08160}., 2017.

\bibitem{lewis2004influence}
M.~Lewis.
\newblock The influence of loyalty programs and short-term promotions on
  customer retention.
\newblock {\em Journal of marketing research}, 41(3):281--292, 2004.

\bibitem{user-exposure}
D.~Liang, L.~Charlin, J.~McInerney, and D.~M. Blei.
\newblock Modeling user exposure in recommendation.
\newblock In {\em Proceedings of the 25th International Conference on World
  Wide Web, {WWW} 2016, Montreal, Canada, April 11 - 15, 2016}, pages 951--961,
  2016.

\bibitem{mutter2014behavioral}
T.~Mutter and D.~Kundisch.
\newblock Behavioral mechanisms prompted by badges: The goal-gradient
  hypothesis.
\newblock In {\em Proceedings of the 35th International Conference on
  Information Systems}, 2014.

\bibitem{Sellke2001}
T.~Sellke, M.~J. Bayarri, and J.~O. Berger.
\newblock Calibration of $\rho$ values for testing precise null hypotheses.
\newblock {\em The American Statistician}, 55(1):62--71, 2001.

\bibitem{wilks1938large}
S.~S. Wilks.
\newblock The large-sample distribution of the likelihood ratio for testing
  composite hypotheses.
\newblock {\em The Annals of Mathematical Statistics}, 9(1):60--62, 1938.

\bibitem{zhang2016social}
J.~Zhang, X.~Kong, and P.~S. Yu.
\newblock {Badge System Analysis and Design}.
\newblock In {\em Proceedings of the 2016 IEEE/ACM International Conference on
  Advances in Social Networks Analysis and Mining}, 2016.

\end{thebibliography}
}

\end{document}